\documentclass[submission,copyright,creativecommons]{eptcs}
\providecommand{\event}{SOS 2007} 
\usepackage{breakurl}             
\usepackage{underscore}           
\usepackage{relsize}
\def \land{\wedge}
\def \lor{\vee}
\def \Equiv{\Leftrightarrow}
\def \IFF{\Equiv}
\newcommand{\eitherp}{\hbox {\rm\bf P}}
\newcommand{\eithert}{\hbox {\rm\bf T}}
\def\start{\hbox{\bf start}}

\newcommand{\nx}{\!\raisebox{-.2ex}{ 
            \mbox{\unitlength=0.9ex
            \begin{picture}(2,2)
            \linethickness{0.06ex}
            \put(1,1){\circle{2}} 
            \end{picture}}}       
            \,}
\newcommand{\ev}{\mathlarger{\diamondsuit}}
\newcommand{\union}{\cup}
\newcommand{\alw}{\raisebox{-.2ex}{
               \mbox{\unitlength=0.9ex
               \begin{picture}(2,2)
               \linethickness{0.06ex}
               \put(0,0){\line(1,0){2}}
               \put(0,2){\line(1,0){2}}
               \put(0,0){\line(0,1){2}}
               \put(2,0){\line(0,1){2}}
               \end{picture}}}
              \,}

\newcommand{\until}{\mathcal{U}}

\newcommand{\ffi}{\varphi}
\newcommand{\imp}{\Rightarrow}

\newcommand{\false}{{\bf\scriptstyle F}}
\newcommand{\true}{{\bf\scriptstyle T}}

\newcommand{\cqd}{\hspace{\fill} \vrule  width  4pt  height  4pt  depth 0pt }

\newcommand{\A}{{\sf A}}
\newcommand{\E}{{\sf E}}

\newtheorem{example}{Example}

\newtheorem{definition}{Definition}
\newenvironment{defn}{\begin{definition}\sl}{\end{definition}}

\newtheorem{theoremaux}{Theorem}
\newenvironment{theorem}{\begin{theoremaux}\sl}{\end{theoremaux}}

\title{On the Expressive Power of the Normal Form for Branching-Time Temporal Logics}
\author{Alexander Bolotov
\institute{University of Westminster\\ London, UK}
\institute{School of Computer Science and Engineering}
\email{A.Bolotov@wmin.ac.uk}
}
\def\titlerunning{On the Expressive Power of the Normal Form for Branching-Time Temporal Logics}
\def\authorrunning{A. Bolotov}

\begin{document}
\maketitle

\begin{abstract}
With the emerging applications that involve complex distributed systems branching-time specifications are specifically important as they reflect dynamic and nondeterministic nature of such applications. We describe the expressive power of a simple yet powerful branching-time specification framework – branching-time normal form, which has been developed as part of clausal resolution for branching-time temporal logics. We show the encoding of B\"{u}chi Tree Automata Tree Automata in the language of the normal form, thus representing, syntactically, tree automata in a high-level way. This enables to translate given problem specifications into the normal form and apply as a verification method a deductive reasoning technique -- the clausal temporal resolution.
\end{abstract}

\section{Introduction}

Automata theoretic methods are fundamental in study of branching-time logics, they are widely used in
the investigations into the expressiveness of branching-time formalisms and their decidability, and are in the centre of the
model-checking techniques \cite{Vardi06automata-theoretictechniques}, being widely used in the framework of formal verification.
With the emerging applications that involve complex distributed systems deployed in heterogeneous environment, for example, 
robotics systems \cite{Fisher-2019-Verification-of-Autonomous-systems}, branching-time specifications are becoming even more important. These specifications reflect the dynamic and non-deterministic nature of such applications and one can envisage the obvious need for efficient techniques to reason about them \cite{Dixon21-clausal-overview}. Tree structures are the underlying models for these specifications and tree automata are well established techniques to reason about tree structures, again, widely used in modern model-checking. On the other hand,  it is useful to have direct methods of deductive reasoning applied to temporal specifications that enable carrying proof. Clausal temporal resolution is one of such techniques \cite{Dixon21-clausal-overview}. The method is based on the concept of the separated normal form  initially developed for resolution based deductive verification in the linear-time framework (see full account of the method in \cite{FDP2001} and an overview of various developments within this framework in \cite{Dixon21-clausal-overview}) and later extended to Computation Tree Logic CTL \cite{Bolotov-phd}, \cite{BF99}, ECTL \cite{BB-jal} and ECTL+ \cite{BB-ECTLplus-journal}. 
These branching-time logics differ in their expressiveness, with CTL allowing only a single temporal operator preceded by a path quantifier, ECTL extending CTL by allowing combinations of temporal operators that express fairness constraints and ECTL+ allowing Boolean combinations of temporal operators and ECTL fairness constraints (but not permitting
their nesting). Yet, the same formalism serves as the normal form for all these logics. For simplicity, here we call this formalism ${\sf BNF}$. The refinement and implementation of the clausal resolution in branching-time setting is given in \cite{ZhangHD09,ZhangHD10}.

In \cite{BFD02} it was shown that the normal form for linear time is as expressive as  B\"{u}chi automata on infinite words \cite{Buc62} while \cite{DixonBF05} defined the translation from an alternating automaton to this normal form and vice versa. In this paper we consider the analogous problem, in the branching-time setup, namely 
we show how ${\sf BNF}$ can represent, {\em syntactically}, B\"{u}chi tree automata in a high-level way. Note that in translating a branching-time problem specification into {\sf BNF},
we actually derive clauses within a fragment of Quantified Computation Tree Logic (EQCTL) \cite{Lar-Quant-CTL-2014}. In particular, formulae
within {\sf BNF} are existentially quantified, moreover, this quantification is external to CTL formulae themselves. Thus, clauses of the {\sf BNF}
satisfy the criteria of the prenex normal form (\cite{Lar-Quant-CTL-2014}), and, translating the given branching-time specification into
{\sf BNF} we stay within this fragment. In order to utilize the normal form as part of proof in the branching-time framework, we effectively skolemize the normal form producing temporal formulae without any quantification.

Having established this relationship between ${\sf BNF}$ and B\"{u}chi tree automata, we justify, theoretically, that ${\sf BNF}$ is applicable to a wider class of branching-time specifications, and a resolution based verification technique can be used as reasoning tool for the obtained specification. 
One important contribution of this paper is that ${\sf BNF}$ specifications obtained from the initial automaton give a good insight into the temporal context, as these specifications are defined with the use of the standard future time temporal operators: $\alw$ (always), $\nx$ (next time), $\ev$ (eventually) and, additionally, path quantifiers $\A$ (on all future paths) and $\E$ (on some future path).
Finally, we note that the result of this paper enables one of the core components of the resolution method - the loop searching (\cite{BD00, Dixon21-clausal-overview}) - to be used to extract (again syntactically) hidden invariants in a wide range of complex temporal specifications.

The rest of the paper is organised as follows. In \S\ref{sec:tree-notation} main technical terms of tree structures
relevant to the paper are introduced.
In \S\ref{ctl-snf} we overview ${\sf BNF}$ outlining its syntax in
\S\ref{snf-langauge} and its semantics, in \S\ref{snf-semantics},
together with an example, \S\ref{snf-example}. In \S\ref{subsection-buchi-automaton} we define
B\"{u}chi tree automata. In \S\ref{buchi-snf}, we show how to translate these automata into
${\sf BNF}$ and establish the correctness while in \S\ref{snf-buchi} the reverse translation from ${\sf BNF}$ into
B\"{u}chi tree automata and its correctness are given. In \S\ref{example} we give an example of the syntactical representation of a small B\"{u}chi tree automaton
in $\sf BNF$. Finally, in \S\ref{discussion}, we provide concluding remarks and discuss future work.

\section{Tree Structure Notation}
\label{sec:tree-notation}

In this section we introduce main concepts of tree structures that are needed for the definition of B\"{u}chi Automata and {\sf BNF}.

\begin{definition}[Paths and Fullpaths of a Tree]
\label{def:tree}
Given a {\bf tree} $T = (N,E)$ such that $N$ is a set of states (nodes) and
$E \subset N \times N$ is a set of edges, with the root $x_0 \in N$, 
a path $\pi_{x_i}$ of a tree $T$ is a (possibly infinite) sequence of nodes
$x_i, x_{i+1}, x_{i+2} \dots$, where for each $j~(i\leq j)$, $(x_j,x_{j+1})\in E$.
A path $\pi_{x_0}$ of a tree with the root $x_0$ is called a \label{fullpaths-intro1} {\bf
fullpath}. We will denote the set of all paths of some tree $T$ by
$\Pi_T$ and the set of all fullpaths by $X_T$.
\end{definition}

Given a tree $T = (N,E)$ the edge relation $E$ is called {\em total} (or connected) if each node $x_i \in N$ belongs to some fullpath.
We will concentrate on such trees where each node is connected with the root.
Now the connectivity \label{connectivity-tree} property of a tree $(N,E)$ can be viewed as the following requirement:
for any state $x_i \in N$ there must be a fullpath $\pi_{x_0}$
such that $x_i \in \pi_{x_0}$, i.e. a path exists which connects this node $x_i$
with the root $x_0$ of a tree.

\begin{definition}[Prefix and Suffix of a path]
\label{def:prefix-suffix}
\qquad Given a path, $\chi_{x_i}$ of a tree $T$ and a node $x_j \in \chi_{x_i}$,
	where $i \leq j$, we call a finite sub-sequence of nodes
	$[x_i,x_j] = x_i, x_{i+1}, \dots x_j$ a {\bf prefix} of a path
	$\chi_{x_i}$ abbreviating it with $Pref(\chi_{x_i},
	[x_i,x_j])$ (or simply as $[x_i,x_j]$ when it is clear which
	path this prefix belongs to) and an infinite sub-sequence of
	nodes $x_j, x_{j+1}, x_{j+2}, \dots $ a {\bf suffix} of a path
	$\chi_{x_i}$ abbreviating it with $Suf(\chi_{x_i}, x_j)$.
\end{definition}

We will utilise the concepts of prefix and suffix in defining the closure properties below.

{\bf Closure properties of CTL models.} When trees are considered
as models for distributed systems, paths through a tree are viewed
as computations. The natural requirements for such models would be
suffix and fusion closures. Following \cite{Emerson90}, the former
means that every suffix of a path is itself a path. The latter
requires that a system, following the prefix of a computation
$\gamma$, at any point $s_j \in \gamma$, is able to follow any
computation $\pi_{s_j}$ originating from $s_j$.

Finally, we require that ``if a system can follow a path
arbitrarily long, then it can be followed forever''
\cite{Emerson90}. This corresponds to limit closure property,
meaning that for any fullpath $\gamma_{s_0}$ and any paths
$\pi_{s_j}, \phi_{s_k}, \dots$ such that $\gamma_{s_0}$ has the
prefix $[s_0, s_j]$, $\pi_{s_j}$ has the prefix $[s_j, s_k]$,
$\phi_{s_k}$ has the prefix $[s_k, s_l]$, etc, and $0 < j < k <
l$, the following holds: 
there exists an infinite path $\alpha_{s_0}$ that is a limit of
the prefixes $[s_0, s_j ], [s_j, s_k], [s_k, s_l], \dots$. Closure properties,
especially, limit closure, are reflected in the formulation of the ${\sf BNF}$, namely, in the
introduction of labels for the clauses of the ${\sf BNF}$.

\begin{definition}[Labelled tree]
\label{Labelled Tree} Given a tree $T = (N,E)$ and a finite
alphabet, $\Sigma$, a $\Sigma$-labelled tree is a structure $(T,L)$ where $L$ is a mapping $N \longrightarrow \Sigma$, which assigns to each node, element
of $N$, some label, element of $\Sigma$.
\end{definition}

\begin{definition}[Branching degree of a node, Branching factor of a tree structure]
\label{def:branching-degree-of-state}
The number, $d$, of\\
immediate successors of a node $x$ in a tree structure is called
the {\em branching degree} of $x$. Thus, $x \cdot k~(1 \leq k \leq d)$
abbreviates the $k-th$ successor of $x$.

Given a set $D =
\{d_1,d_2, \dots\}$, of the branching degrees of the nodes of a
tree structure, the maximal $d_i~(1\leq i)$ is called the {\em
branching factor} of this tree structure. 
A tree structure with its branching factor $d$ is called a $d$-ary tree structure.
\end{definition}

We assume that underlying tree
models are of at most countable branching factor. However, following
(\cite{Emerson90}, page 1011) trees with arbitrary, even
uncountable, branching, `as far as our branching temporal logic
are concerned, are indistinguishable from trees with finite, even
bounded, branching'. This makes tree automata applicable in the branching-time framework.
We will also use this result to justify the labelling of {\sf BNF} clauses.

\begin{definition}[$\mathbf\omega$-tree]
\label{omega-tree-def}
An {\bf $\mathbf\omega$-tree} $(N,E^{\omega})$ is a tree,
which satisfies the following conditions.
\begin{enumerate}
\item A tree is of at most countable branching.
\item The relation $E$ is {\em serial}, i.e.\hspace{-0.5ex} every state $s_i$
must have at least one successor state.
\item $E$ induces the natural ordering
$\leq$: if $(s_i,s_j)\in E^{\omega}$
then $i \leq j$, where $\leq$ orders the set of natural numbers
$\omega = \{0,1,2,\dots\}$.
\end{enumerate}
\end{definition}

Now, following \cite{ES84}, given that a CTL model structure
${\cal M}$ (see \S\ref{snf-semantics} for the definition of a model structure) has its branching factor at most $d$, there exists a
$d$-ary tree canonical model ${\cal M}'$ such that for any formula
$A$, ${\cal M}$ satisfies $A$ if, and only if, ${\cal M}'$
satisfies $A$. Informally, a canonical model is an unwinding of
an arbitrary model ${\cal M}$ into an infinite tree $T$
\cite{ES84}. One of the essential properties of canonical models, which we will utilise here, 
is that the number of successors for every state is canonicalised by $d$.   

\section{Normal form used for the clausal resolution for CTL-type logics}
\label{ctl-snf}
\vspace{1ex}

Normal form {\sf BNF} which we are considering, is a formalism that has been developed as part of
the clausal resolution method developed initially for linear-time temporal logic \cite{Fi97, FDP2001} and then defined for branching-time temporal logics, CTL \cite{BF99}
and its extensions, ECTL \cite{BB-jal} and ${\sf ECTL}^+$ \cite{BB-ECTLplus-journal}.
As one would expect from clausal resolution, formulae of a given logic are first translated into normal form, which is a collection of clauses, to which a resolution method is applied. The idea behind the resolution procedure in temporal context is to extract for some given formula $A$ a set of clauses that capture three types of `knowledge' about $A$ - what is happening at the beginning of the computation, what are the `steps' of the computation, and what are the eventualities to be fulfilled during the computation. 

For a formula $A$ of a CTL-type branching-time logic, we will abbreviate its clausal normal form as ${\sf BNF}(A)$. Note that the standard formulation of CTL-type logics is based upon classical connectives, future time temporal operators 
$\nx$ and $\until$ (until) (which is sufficient to introduce $\alw$ and $\ev$) 
and path quantifiers $\A$ and $\E$. The {\sf BNF}, however, is formulated using $\nx$, $\alw$ and $\ev$, and $\start$ (which is only true at the beginning of the computation) but not utilising the $\until$ operator as it is removed during the translation to the normal form based on its fixpoint definition \cite{BF99}.

\subsection{Language of BNF}
\label{snf-langauge}

In the language for BNF we utilise 
\begin{itemize}
\item classical connectives: implication ($\imp$), negation ($\lnot$), disjunction ($\lor$), and conjunction ($\land$); 
\item classically defined constants true ($\true$) and false ($\false$); 
\item temporal operators `at the initial moment of time' ($\start$), eventually ($\ev$), always ($\alw$), next time ($\nx$), and, 
\item path quantifiers: on all future paths ($\A$) and on some future path ($\E$).
\end{itemize}

In the rest of the paper we will use the following notation: \eithert{} abbreviates any unary
${\sf BNF}$ temporal operator and \eitherp{} either of path quantifiers; any formula of the type \eitherp\eithert{} is called {\em a basic} ${\sf BNF}$ {\em modality}, and 
a {\em literal} is a proposition or its negation.

{\bf Indices.} The language for indices is based on the set of
terms $\{ IND \}  =  \{ {\sf f},~{\sf g},~{\sf h}, \dots \}$, where ${\sf f,~g,~h}\dots$ denote constants. Note that indices play essential role in
the formulation of ${\sf BNF}$ as they help identifying a specific path context for given formulae. We use indices to label all formulae of ${\sf BNF}$ that contain the basic modality $\E\nx$ or $\E \ev $. Specifically, the modality $\E\nx$ is associated with ${\sf BNF}$
step clauses (see below) and, thus, if $\E \nx A_{\sf g}$ is true at some current state $s_i$ then $A$ holds at the successor state of $s_i$ along the path associated with the `direction' ${\sf g}$ - speaking informally, we only take `one step' in direction $g$ from $s_i$.
The modality $\E\ev$ is associated with evaluating eventualities over a longer period of time, 
and thus if $\E\ev p_{\sf g}$ is true at some state $s_i$ then $p$ is true at some state $s_k$ along the path which goes from $s_i$ by making at $s_i$, and every subsequent successor state, a `mini-step' along the `direction' ${\sf g}$. This corresponds to the limit closure of the concatenation of these `mini-steps' in directions ${\sf g}$ and hence the existence of such a path is always guaranteed.


\begin{definition}[Branching Normal Form]
\label{def:snfctl} Given $Prop$, a set of atomic propositions, and $IND$, a countable set of indices,
${\sf BNF}$ has the structure
$$\A\alw\displaystyle{\left[\bigwedge_i C_i\right]}$$ where each of the
{\em clauses\/} $C_i$ is defined as below and
each $\alpha_i$, $\beta_j$ or $\gamma$ is a literal, $\true$ or
$\false$ and ${\sf ind} \in IND$ is some index.
\end{definition}

\noindent
\begin{center}
$
\begin{array}{rcll} \start &\imp &
\displaystyle\bigvee_{j=1}^k \beta_j
        \qquad& \hspace{-3ex}\hbox{an~Initial~Clause}\\[1.5ex]
\displaystyle\bigwedge_{i=1}^l \alpha_i
    &\imp &\A\nx \displaystyle{\left[\bigvee_{j=1}^k \beta_j
    \right]}
    & \hspace{-3ex}\hbox{an~\A~step~clause}\\
\displaystyle\bigwedge_{i=1}^l \alpha_i
    &\imp &\E\nx \displaystyle{\left[\bigvee_{j=1}^k{\beta_j}\right]_{\sf ind}}& \hspace{-3ex}\hbox{a~\E~ step~clause}\\[1.5ex]
\displaystyle\bigwedge_{i=1}^l \alpha_i
        &\imp & \A\ev \gamma
        &\hspace{-3ex} \hbox{an~\A~sometime~clause}\\[1.5ex]
\displaystyle\bigwedge_{i=1}^l \alpha_i
        &\imp & \E\ev\gamma_{\sf ind}
        & \hspace{-3ex}\hbox{a~\E~sometime~clause}\\
\end{array}
$
\end{center}
\cqd

The intuition behind this formulation is that initial clauses provide the initial conditions for the computation while each of the step
clauses represents a constraint upon the future behaviour of the
formula, given the current conjunction of literals.  
Note that the $\A\alw$ constraint is only utilised BNF as the one surrounding the conjunction of clauses which gives the clauses the `universal' interpretation by propagating information they represent to each state along each path of the tree structure. 
At the same time, the $\E\alw$ constraint is not utilised in the BNF, although, obviously it can be introduced based on $\E\alw\alpha =_{\sf def} \E\lnot\ev\lnot\alpha$.
Finally note that in the eventuality clauses the argument of the $\ev$ operator is a literal. This requirement is due to the potential application of the clausal resolution method to the specifications written in BNF. The clausal resolution method resolves (if possible) an eventuality, $\ev l$ in the scope of a path quantifier with the loop in $l$ which is defined on all or some dedicated paths, see details in \cite{Bolotov-phd, BB-jal}.

\subsection{Interpretation of {\sf BNF}}
\label{snf-semantics}

For the interpretation of ${\sf BNF}$ clauses, utilising the notation of \cite{ZhangHD09},
we introduce an indexed tree-like model. Given $IND$ is a countable set of
indices, $S$ is a set of states, $E\subseteq S\times S$ is a total binary relation over
$S$, and $L$ is an interpretation function $S \longrightarrow 2^{Prop}$, which maps a state $s_i\in S$ to the set of atomic propositions that are true at $s_i$.
Then an indexed model structure ${\cal M} = \langle S,R,L, {\overrightarrow {\sf ind}}, s_0 \rangle$ where
$s_0\in S$, and ${\overrightarrow {\sf ind}} \subseteq R$ such that it is the argument of the  mapping of every index ${\sf ind} \in IND$ to the successor function ${\overrightarrow {\sf ind}}$ such that
${\overrightarrow {\sf ind}}$ is a total functional relation on $S$



It is easy to see that the underlying tree model above is an $\omega$-tree
satisfying the conditions of Definition \ref{omega-tree-def}.

A state $s_j \in S$ is an ${\sf ind}$-successor state of state $s_i \in S \IFF
(s_i, s_j) \in {\overrightarrow {\sf ind}}$.
An infinite path $\chi_{s_i}^{\sf ind}$ is an infinite sequence of states $s_i, s_{i+1}, s_{i+2}, \dots$
such that for every $j~(i \leq j)$, we have that $(s_j , s_{j+1}) \in {\overrightarrow {\sf ind}}$.

Below, we define the relation `$\models$', omitting cases for Booleans, $\true$ and $\false$. Also recall that 
the basic modality $\E\alw$ is not used in the BNF while $\A\alw$ only appears as an outer modality preceeding the conjunction of the BNF clauses.

$
\hspace{-2ex}
\begin{array}{rclll}
\langle{\cal M}, s_i\rangle        &\models& \start &\IFF& i=0\\
\langle {\cal M}, s_i\rangle        &\models& \A \nx B    &\IFF& for~each~ {\sf ind} \in IND
                                    ~and ~each~s' \in S, if ~(s_i,s')\in \overrightarrow {\sf ind} 
                                    ~then ~ \langle {\cal M}, s'\rangle \models B\\
\langle {\cal M}, s_i\rangle        &\models& \E \nx B_{\sf ind}    &\IFF& there~exists~s' \in S, such~ that ~(s_i,s')\in {\overrightarrow {\sf ind}} ~and ~ \langle {\cal M}, s'\rangle \models B\\
\langle {\cal M}, s_i\rangle        &\models& \A \alw B    &\IFF& for~each~ \chi_{s_i}and~s_j\in \chi_{s_i},
                                    ~{\sf if}~ (i \leq j)~then~ \langle {\cal M}, s_j\rangle \models B\\
\langle {\cal M}, s_i\rangle        &\models& \A \ev B    &\IFF& for~each~ \chi_{s_i},
                                    there~exists~s_j\in S,~such~that~ (i \leq j)
                                    and~ \langle {\cal M}, s_j\rangle \models B\\
\langle {\cal M}, s_i\rangle        &\models& \E \ev B_{\sf ind} &\IFF& ~there~exist~\chi_{s_i}^{\sf ind} and~s_j\in \chi_{s_i}^{\sf ind}, ~such~that~i \leq j
                                    ~ \langle {\cal M}, s_j\rangle \models B\\
\end{array}
$

\begin{defn}[Satisfiability, Validity] If $\cal C$ is in ${\sf BNF}$ then
$\cal C$ is satisfiable if, and only if, there exists a model ${\cal M}$ such
that $\langle {\cal M},s_0\rangle\models \cal C$.
$\cal C$ is
valid if, and only if, it is satisfied in every possible model.
\end{defn}

\subsection{${\sf BNF}$ examples.}
\label{snf-example}

Here we give an example of the normal form and informal interpretation.
Noting that an initial ${\sf BNF}$ clause, $\start \imp F$, is understood as ``$F$
is satisfied at the initial state of some model ${\cal M}$''and any
other ${\sf BNF}$ clause is interpreted taking also into account
that it occurs in the scope of $\A\alw$ let us consider a
clause $\A\alw(x \imp \E\nx q_{\sf
ind})$. It is understood as
``{\em for any fullpath $\chi$ and any state $s_i\in \chi~(0 \leq
i)$, if $x$ is satisfied at a state $s_i$ then $q$ must be
satisfied at the moment, next to $s_i$, along some path associated
with ${\sf ind}$ which departs from $s_i$}''.

The clause $\A\alw(x \imp \E\ev p_{\sf ind})$ has the following meaning
``{\em for any fullpath $\chi$ and any state $s_i\in \chi~(0 \leq
i)$, if $x$ is satisfied at a state $s_i$ then $p$ must be
satisfied at some state, say $s_j~(i\leq j)$, along some path
$\alpha_{s_i}$ associated with the limit closure\footnote{Observe that limit closure has the properties of the reflexive transitive closure.} of ${\sf
ind}$ which departs from $s_i$}'.



\section{B\"{u}chi Tree automata}
\label{subsection-buchi-automaton}

\begin{definition}
\label{buchi-automaton-def}
A B\"{u}chi automaton, $\cal B$, on an infinite tree, $T$, is a tuple
${\cal B} = \langle \Sigma,~D,~S,\delta,F_0,~F_B\rangle$ where:
$\Sigma$ is a finite alphabet;
$D$ is a finite set of branching degrees; 
$S= \{s_0,s_1,\dots s_k\}$ is a finite set of states;
$F_0 \subseteq S$ is a set of initial states;
$\delta$
is a non-deterministic transition function satisfying;
$\delta(s,\sigma,d) \subseteq S^d$, for every $s\in S,\sigma\in \Sigma$ and $d\in D$, and
$F_B\subseteq S$ is a set of accepting states. 
\end{definition}

The transition function $\delta(s,\sigma, d)$ is the set of all tuples of states to which the automaton may evolve from state $s\in S$ of the arity $d\in D$ when it reads the symbol $\sigma \in \Sigma$. 
 
 A {\em run}, $\tau_{\cal B}: T \longrightarrow S$, of a B\"{u}chi tree automaton ${\cal B}$
over the input $\Sigma$-labelled tree $(T ,L)$ is an $S$ labelled tree such that the root is labelled by a member of $F_0$
and the transitions conform with the transition function $\delta$.
Namely, visiting a state $s_i\in S$ with the branching degree $d$ and reading $\sigma\in \Sigma$, an automaton makes a non-deterministic choice of
a tuple $s_0, \dots s_d \in \delta(s,\sigma,d)$, $1 \leq d \leq k$, makes $d$ copies of itself and moves to the node $s_i\cdot s_j~(1\leq j \leq d)$.
\vspace{1.5ex}

A run, $\tau_{\cal B}$, is {\em successful} if for every infinite branch of $\tau_{\cal B}$,
there is an accepting state $s \in F_B$ that occurs infinitely often in this branch.
An automaton ${\cal B}$ accepts the infinite tree $T$
(in other terms, the language recognised by ${\cal B}$ is {\em not empty})
if it has a successful run $\tau_{\cal B}$.

\section{From B\"{u}chi Tree automata to ${\sf BNF}$}
\label{buchi-snf}


Here, given a B\"{u}chi tree automaton ${\cal B}$, we construct its
characteristic formula, ${\sf BNF}_{\cal B}$, as a set of ${\sf BNF}$
clauses, following the main stages similar to those in
\cite{BFD02}: encoding of the set of the initial states, representing the run of
the automaton, the labelling, and the acceptance condition. 

Let ${\cal B} = \langle \Sigma,~D,~S,\delta,F_0,~F_B\rangle$
be a B\"{u}chi tree automaton with the states labelled
as follows. For every proposition $p\in Prop$, for $\sigma \in \Sigma$, given
$s_i\cdot s_k \in \delta(s_i, \sigma, d)$ (for $1\leq k \leq d$), if $p \in \sigma$ then $p\in L(s_i\cdot s_k)$ else
if $p\not\in \sigma$ then $\lnot p \in L(s_i\cdot s_k)$.
In our translation we will explicitly encode this labelling.
Now, 
we introduce formulae
(\ref{init-snf-direct-translation})--(\ref{acceptance-formula-snf-direct}),
which represent the main stages of the translation. 
Note that to simplify reading we will omit writing the outer $\A\alw$, 
and will write a set of ${\sf BNF}_{\cal B}$ clauses rather than their
conjunction; in some cases we will alos give the formulation of the components
of the translation not in the exact
form of ${\sf BNF}_{\cal B}$ clauses to make the idea behind the translation
more explicit. Each of these cases, where we need further
simple manipulations to obtain the required BNF form, will
be marked with $\{\star\}$.

Let $q_0,\dots, q_l$ be new propositions of our ${\sf BNF}$ language
such that $q_i~(0\leq i\leq l)$ encodes the state $s_i \in S$ of the automaton.
Given that $q_0, \dots, q_m,~(0 \leq m \leq l)$ encode the initial states, $F_0$, of the automaton, we
specify $F_0$ by the following ${\sf BNF}$:
\begin{equation}
\label{init-snf-direct-translation}
{{\sf BNF}}_{init_{\cal B}}:
\begin{array}[t]{rrcl}
(\ref{init-snf-direct-translation}.1)&\start &\imp&  q_0 \lor \dots \lor q_m\\
(\ref{init-snf-direct-translation}.2)&\start &\imp&  \lnot q_i \lor \displaystyle\bigwedge_{i\not= j}(\lnot {q_j})\quad \{\star\}\\
\end{array}
\end{equation}
\noindent where $0 \leq i \leq m,~0 \leq j \leq m$. From  (\ref{init-snf-direct-translation}) it follows that the automaton can be at only one initial state at the first moment of time. Note that constraint (\ref{init-snf-direct-translation}.2) of the ${\sf BNF}_{init_{\cal B}}$ marked with `$\star$' should be further translated to the form required by the BNF.

\noindent Next, the transition function of the automaton is represented as follows.

\noindent Given the automaton ${\cal B}$ with the set of branching degrees $D = \{1, \dots, d\}$, we associate with each element of the latter a set of new indices $IND = ind_1, \dots ind_d$ used to label ${\sf BNF}$ clauses. Thus, when the automaton makes its $d$ copies visiting a state $s_i\in S$, with the branching factor $d$,
with each such successor node $s_n$ of $s_i$ we associate an index $ind_j, ~ 1 \leq j \leq d$.

\begin{equation}
\label{run-snf-direct-translation}
\begin{array}[t]{ll}
{{\sf BNF}}_{tran_{\cal B}}:
&
\quad (\ref{run-snf-direct-translation}.1) q_i \imp \E\nx {q_n}_{ind_j}\\
\end{array}
\end{equation}

\noindent Thus each of $d$ clauses in (\ref{run-snf-direct-translation}) reflects a successor node of $s_i$ along the path labeled by $ind_j$. Note that unlike B\"{u}chi word automaton which can only be at one state at any particular time, a tree automaton makes copies choosing a corresponding tuple of states. This is exactly what we have represented above. Thus, for each $q_i$ of branching degree $d$ there are exactly $d$ clauses of the form (\ref{run-snf-direct-translation}).

Next, we represent the unique labelling of the states of the automaton by the following
set of clauses, constructed for every $q_i$ and every $x_i\in L(s_i)$.
\begin{equation}
\label{labels-snf-direct-translation}
\begin{array}[t]{ll}
{BNF}_{lab_{\cal B}}:&
\begin{array}[t]{rlrclr}
&(\ref{labels-snf-direct-translation}.1)& \start &\imp& \lnot q_i \lor 
\displaystyle{\left[\left(\bigwedge_{x_i\in L(s_i)} x_i\right) \land \left(\bigwedge_{x_j\not\in L(s_i)} \lnot x_j\right)\right]} &\{\star \}\\
\\
&(\ref{labels-snf-direct-translation}.2)& \true  &\imp& \A \nx (\lnot q_i \lor v)\\
\\
&((\ref{labels-snf-direct-translation}.3)& v &\imp& \displaystyle{\left[\left(\bigwedge_{x_i\in L(s_i)} x_i\right) \land \left(\bigwedge_{x_j\not\in L(s_i)} \lnot x_j\right)\right]}  & \{\star \}
\end{array}
\end{array}
\end{equation}
The B\"{u}chi acceptance condition
is given by the following set of ${\sf BNF}$ clauses constructred for each $ind_i \in IND$.
\begin{equation}
\label{acceptance-formula-snf-direct}
\begin{array}[t]{ll}
{BNF}_{acc_{\cal B}}:
&
\begin{array}[t]{rlrcl}
&(\ref{acceptance-formula-snf-direct}.1)& \start &\imp& y\\
&(\ref{acceptance-formula-snf-direct}.2)&    y   &\imp& \A\alw u\quad\quad \{\star \}\\
&(\ref{acceptance-formula-snf-direct}.3)&    u   &\imp& \E\ev l_{ind_i}\\
&(\ref{acceptance-formula-snf-direct}.4)&    l   &\imp& \E\nx w_{ind)i}\\
&(\ref{acceptance-formula-snf-direct}.5)&    w   &\imp& \E\ev l_{ind)i}\\
&(\ref{acceptance-formula-snf-direct}.6)& \start &\imp& \lnot l \lor q_n \lor\dots \lor q_r\\
&(\ref{acceptance-formula-snf-direct}.7)& \true &\imp& \A\nx(\lnot l \lor q_n \lor \dots \lor q_r)\\
\end{array}
\end{array}
\end{equation}
where 
\begin{itemize}
\item $q_n\dots q_r$ encode the accepting states of the
automaton and $y,~w,~l$ and $z$ are new propositions. This condition ensures that
every path of the run hits an accepting state from $F_B$ infinitely often. The use of the indices
guarantees that we are staying in the `context of a chosen path' when verifying the acceptance condition. 
\item Constraint 2 of the ${BNF}_{acc_{\cal B}}$ marked with `$\star$' is the abbreviation
of the `loop' in $u$. This loop in the BNF is represented by the following two clauses
$({\sf i})~y  \imp (z\land u), ({\sf ii})~z\imp \A\nx (u \land z)$ (recall that all clauses are in the scope of the $\A\alw$): from $({\sf i})$ we know that $u$ is true at the state where $y$ is true, and also that $z$ is also true at that state, while from $({\sf ii})$ we derive that the every successor state should satisfy both $u$ and $z$. This second clause $({\sf i})$ ensures the recurrent presence of $z$ in every subsequent state along every path, which, in turn, ensures that each subsequent state along every path also satisfies $u$.\\
\end{itemize}

Finally, let ${\sf BNF}_{init_{\cal B}}'$, ~${\sf BNF}_{tran_{\cal B}}'$,~ ${\sf BNF}_{lab_{\cal B}}$
and ${\sf BNF}_{acc^B_{\cal B}}$ be obtained from
${\sf BNF}_{init_{\cal B}}$, ${\sf BNF}_{tran_{\cal B}}$,\\
~ ${\sf BNF}_{lab_{\cal B}}$ and  ${\sf BNF}_{acc^B_{\cal B}}$, respectively, by translating their
components into the required form of ${\sf BNF}$ clauses. 
Now, a B\"{u}chi tree automaton $\cal B$ is characterized by the
following ${\sf BNF}$ expression known as a {\em characteristic clause set} and
abbreviated by ${\sf BNF}_{\cal B}$:
\begin{equation}
\label{characteristic-clause-set}
{\sf BNF}_{{init_{\cal B}}}' \land {\sf BNF}_{{tran_{\cal B}}}' \land{\sf BNF}_{{lab_{\cal B}}}
\land {\sf BNF}_{{acc^B_{\cal B}}}'
\end{equation}

\subsection{Correctness}

\begin{theorem}
\label{theorem-snfctl-satisfiable-iff-accepting-run}
Given a B\"{u}chi tree automaton ${\cal B}$, we can construct a
characteristic clause set, ${\sf BNF}_{\cal B}$, such that ${\cal B}$
has an accepting run, $\tau_{\cal B}$ (over an infinite tree $T$),
if and only if, ${\sf BNF}_{\cal B}$ is satisfiable.
\end{theorem}

{\bf Proof.}

\noindent (I) Left to right direction. The proof effectively follows the labelling chosen for ${\sf BNF}$ clauses described above.
Given a B\"{u}chi tree automaton, ${\cal B} = \langle \Sigma,~D,~S,\delta,F_0,~F_B\rangle$, on an infinite tree,
recall that its accepting run $\tau_{\cal B}$ is an $S$-labelled tree $(T ,L)$ such that its root is labelled by a member of $F_0$
and the transitions conform with the transition function $\delta$.

{\em STEP 1.} First, let $|IND| = d$, where $d\in D$ is the largest branching factor, and let the states of $S$ be obtained from the corresponding nodes of $T$ provided the states of a B\"{u}chi automaton are labelled as above, namely, such that for every proposition $p\in Prop$, for $\sigma \in \Sigma$, given
$s_i\cdot s_k \in \delta(s_i, \sigma, d)$ (for $1\leq k \leq d$), if $p \in \sigma$ then $p\in L(s_i\cdot s_k)$ else
if $p\not\in \sigma$ then $\lnot p \in L(s_i\cdot s_k)$. This labelling guarantees that the mapping $\overrightarrow {\sf ind}$ establishes the desired order over
the states of $S$ so every state $x\cdot k$ with the degree $d$ ($1\leq k \leq d$) of the tree model, is identified with the $S$-label of the node $x\cdot k$ of the run $\tau_{\cal A}$. Thus, the way how the labelling chosen for the ${\sf BNF}$ clauses guarantees that this tree becomes an underlying tree structure for the model ${\cal M} = \langle S,R,L, {\overrightarrow {\sf ind}}, s_0 \rangle$ such that for every $p_k \in L(x\cdot k)$, $({\cal M},s\cdot k) \models p_k$.

{\em STEP 2.} Let a model ${\cal M}' = \langle S',R,L', {\overrightarrow {\sf ind}}, s_0' \rangle$ be the same
as ${\cal M}$ except for the interpretation of the new propositions $q_0,q_a, \dots q_b, q_n, q_r, y,l, u, v, w$ and $z_1 \dots z_m$
which we chose to satisfy at the appropriate states of $S'$. For example, $\langle{\cal M}',s_0 \rangle \models y$ and for
every $s_n\not= s_0$, $\langle{\cal M}',s_n \rangle \not\models y$. Updating this way the model ${\cal M}$ into ${\cal M}'$ we guarantee
that each component of the characteristic clause set is satisfied in ${\cal M'}$

\noindent (II). Right to left direction. We prove this by contradiction, i.e. assuming that the given B\"{u}chi tree automaton, $${\cal B} = \langle \Sigma,~D,~S,\delta,F_0,~F_B\rangle$$ on an infinite tree, does not have an accepting run, we show that we cannot build a model for the characteristic clause set, ${\sf BNF}_{\cal B}$, for ${\cal B}$. The emptiness of the automaton would mean that the acceptance condition is violated.
Thus, as there is no succefull run of the automaton, it should have an infinite branch $\chi$, such that 
there is no accepting state $s_i \in F_B$ which occurs in ${\chi}$ infinitely often.
According to the construction of the intermediate model structure for every $p_k \in L(x\cdot k)$, $({\cal M}, s\cdot k) \models p_k$.
Any model ${\cal M}' = \langle T',\leq, I'\rangle$ which agrees with ${\cal M}$ everywhere except for the interpretation of the new
propositions appeared in the characteristic clause set, does not satisfy the latter.
Indeed, since $q_n,\dots,q_r$ are labels of the accepting states, and none of the accepting states occurs in $\chi$ infinitely often, for every $q_j~(n\leq j\leq r)$, for any ${\cal M}'$ we have that $\lnot q_j$ becomes eventually always true along $\chi$, and hence, in ${\cal M}'$ the conjunction $\lnot q_n \land \dots \land \lnot q_r$ becomes always true from some point on along this path $\chi$. 
This contradicts the satisfiability conditions for ${\sf BNF}_{{{\cal B}}}'$: because the conjunction $\lnot q_n \land \dots \land \lnot q_r$ becomes always true from some point on, say $s_m\in S$ along $\chi$, given also (\ref{acceptance-formula-snf-direct}.6) and (\ref{acceptance-formula-snf-direct}.7) we must have a loop in $\lnot l$, from $s_m$, i.e. $\alw\lnot l$ becomes true along $\chi$ from $s_m$. As (\ref{acceptance-formula-snf-direct}.3) and (\ref{acceptance-formula-snf-direct}.5) are constructed for 
each $ind_i \in IND$, we would have a contradiction between this loop in $\lnot l$ along $\chi$ and the request to fulfil the eventuality $l$ along this path. 

\cqd

\section{From ${\sf BNF}$ to B\"{u}chi Tree automata}
\label{snf-buchi}

In this section we show how to effectively construct a B\"{u}chi Tree automaton from a given set ${\cal C}$ of ${\sf BNF}$ clauses. First, let us distinguish among
${\sf BNF}$ clauses the initial clauses and the global (step and sometime) clauses. The ideology is as follows.
First we apply the augmentation technique developed for the clausal resolution for CTL \cite{Bolotov-phd} deriving ${\cal C}_{Aug}$ an augmented ${\sf BNF}$.
Then we show how to construct a model for ${\cal C}_{Aug}$. This technique involves a construction of a tableau as a labelled finite directed graph.
The states of the graph are labelled by the sets of subformulae of ${\cal C}_{Aug}$. Let $Prop({\cal C}_{Aug})$ be a set of all (different) propositions that occur within the clauses of ${\cal C}_{Aug}$. The important features of the underlying transition function for the construction of the tableau is that the formula $\A\alw \phi$, where $\phi$ is a conjunction of all global clauses of ${\cal C}_{Aug}$, occurs within each state.
Thus, global clauses play the role of a guide for the transitions. A transition from a state $s_i$ to a state $s_j$ is provided if a label for $s_j$ is consistent.

Once a tableau ${\cal G}_{{\cal C}_{Aug}}$ is constructed, labels of some states
might contain formulae of the type $\eitherp\ev l$. Thus, we check if such an eventuality
is satisfied. During this procedure we delete those
states (and their successors) of a graph  ${\cal G}_{{\cal C}_{Aug}}$ which contain
unsatisfied eventualities. As some states of a graph now might be
without any successor, we will delete such states. The resulting graph, ${\cal R}$,
is called a {\em reduced tableau}.

It has been shown that the set of ${\sf BNF}$ clauses is unsatisfiable,
if, and only if, its reduced tableau is empty \cite{Bolotov-phd}.

\begin{definition}[\bf Augmentation]
\label{def-tres-augmentation}

Given a set of ${\sf BNF}$ clauses, ${\cal C}$, we construct an {\bf augmented} set ${\cal C}_{Aug}$ as follows.

\begin{enumerate}
\item Create a list $\{EVEN\} = \E\ev l_1, \E\ev
l_2, \dots,$ $\dots\E\ev {l_n}, \A\ev {l_{n+1}},
\A\ev {l_{n+2}}, \dots , \A\ev l_{m}~
(0 \leq n \leq m)$ of all different eventualities contained in ${\cal C}$.
\item To keep track of the eventualities, create a list $W = w_1, w_2, \dots, w_m$ of new propositions (that
do not occur within clauses of ${\cal C}$) such that each $w_i\in W$ is associated
with the $i$-th eventuality within $EVEN$.
\item For every sometime clause $C \imp \eitherp\ev {l_i}_{\sf ind_i}$,
to guarantee the correspondence between $w_i$ and $l_i$, add the
corresponding set of formulae, defined below.
\begin{itemize}
\item[3a] If $\eitherp\ev l_i = \A\ev l_i$, then
add to the set of ${\sf BNF}$ clauses ${\cal C}$, the following formulae:

\begin{equation}
\label{eq:augmentation-3a}
\begin{array}{rcl}
\start  & \imp & \lnot C \lor l_i \lor w_i\\
w_i     & \imp & \A\nx (l_i \lor w_i)\\
\true   & \imp & \A\nx (\lnot C \lor l_i \lor w_i).
\end{array}
\end{equation}

\item[3b] If $\eitherp\ev l_i = \E\ev {l_i}_{\sf ind}$, then
add to the set of ${\sf BNF}$ clauses ${\cal C}$, the following formulae:
\begin{equation}
\label{eq:augmentation-3b}
\begin{array}{rcl}
\start  & \imp & \lnot C  \lor l_i \lor w_i\\
w_i     & \imp & \E\nx (l_i \lor w_i)_{\sf ind}\\
\true   & \imp & \E\nx (\lnot C \lor l_i \lor w_i)_{\sf ind}
\end{array}
\end{equation}

\end{itemize}
\end{enumerate}
\end{definition}

Let ${\cal C}_{Aug}$ be an augmented set of ${\sf BNF}$ clauses. Abbreviating by $In$ the conjunction of the right hand sides of
the initial clauses of ${\cal C}_{Aug}$ and by $\phi$ the conjunction of its step and
eventuality clauses, we can represent a set, ${\cal C}_{Aug}$
as a formula $In\land \A\alw \phi$. It is easy to show from the BNF semantics that the following holds:
$\langle {\cal M}, s_0 \rangle \models$ $\cal C$ $\IFF \langle {\cal M},s_0 \rangle \models
In\land \A\alw \phi$.

Let an {\em elementary} formula be either a literal, or has its main
connective as $\eitherp\nx$. Each {\em non-elementary} \label{ctl-tableau-elementary-formula} formula is further classified as either
a conjunctive, $\alpha$-formula, or a disjunctive, $\beta$-formula.

A basic ${\sf BNF}$ modality now is qualified according to its fixpoint definition (in the equations below $\mu$ and $\nu$ stand for `minimal fixpoint' and `maximal fixpoint' operators, respectively)

\begin{equation}
\label{max-fixpoints}
\begin{array}{l}
\A\alw\ffi =  \nu \rho (\ffi \wedge \A\nx \rho)\\
\E\ev\ffi  =  \mu \rho (\ffi \vee \E\nx \rho)\\  
\A\ev\ffi  =  \mu \rho (\ffi \vee \A\nx \rho) 
\end{array}
\end{equation}

The only ${\sf BNF}$ modality that occurs within ${\sf BNF}$ clauses as
a maximal fixpoint is $\A\alw \phi$ and in our representation of a set of ${\sf BNF}$ clauses
as $In \land \A\alw \phi$, it has the only occurrence as the main connective in $\A\alw \phi$, hence, the following
$\alpha$-expansion \label{alpha-rules} rules will be appropriate:

\begin{equation}
\label{eq:alpha-rules}
\begin{array}{l|l|l}
\alpha& \alpha_1 &\alpha_2\\
\hline
B\land C & B &C \\
\A\alw B & B &\A\nx\A\alw B\\
\lnot (B \lor C) & \lnot B & \lnot C \\
\end{array}
\end{equation}

${\sf BNF}$ modalities understood as a minimal fixpoints are
$\A\ev \phi$ and $\E\ev \phi$,
hence, the following $\beta$-expansion \label{betha-rules} rules will be appropriate:

\begin{equation}
\label{eq:beta-rules}
\begin{array}{l|l|l}
\beta& \beta_1 &\beta_2\\
\hline
B \imp C & \lnot B &C \\

B \lor C & B &C \\

\A\ev B &  B  &\A\nx\A\ev B\\

\E\ev B_{\sf ind} & B &\E\nx\E\ev B_{\sf ind}\\
\end{array}
\end{equation}

\noindent Here the last two constraints must be further transformed into the appropriate structure of ${\sf BNF}$ clauses, however, the current representation
is preferable as it illustrates the intuition.
Let $Even_{{\cal C}_{Aug}}$ be a list of eventualities as defined in Definition \ref{def-tres-augmentation}
and let $Prop_{{\cal C}_{Aug}}$ be a set of all (different) propositions that occur within the clauses of ${\cal C}_{Aug}$. By an evaluation of a proposition $p_i \in Prop_{{\cal C}_{Aug}}$ we understand the function $Prop_{{\cal C}_{Aug}} \longrightarrow {0,1}$. Now, let 
$Val(Prop_{{\cal C}_{Aug}})$ be a set of all possible evaluations of the elements of
the $Prop_{{\cal C}_{Aug}}$. Finally, let $D = {\sf ind_1}, {\sf ind_2} \dots {\sf ind_k}$ be a list
of all different indices not of the form {\sf ind} which occur in ${\cal C}$.

We adapt the notion of the generalized Fischer-Ladner closure \cite{FL79} introduced for CTL formulae
in \cite{EH85} for our case of ${\sf BNF}$.

\begin{definition}[\bf Generalized Fischer-Ladner closure
for  ${\sf BNF}$]
\label{ctl-fl-closure}
\end{definition}
{\em Let ${\cal C}$ be a set of ${\sf BNF}$ clauses, let $In \land \A\alw \phi$ be its
equivalent formula, where `$In$' abbreviates the
conjunction of the right hand sides of the initial clauses of ${\cal C}$ and
$\phi$ abbreviates the conjunction of the global clauses within ${\cal C}$.
Then the least set of formulae which contains ${\cal C}$ and satisfies the
conditions below is the generalised Fischer-Ladner closure of ${\cal C}$, abbreviated by $GFL({\cal C})$.}
\begin{itemize}
\item[]
\begin{itemize}
\item[(GFL1)] $In \land \A\alw \phi$ is an element of $GFL({\cal C})$.
\item[(GFL2)] If $B$ is an element of $GFL({\cal C})$ then any subformula of $B$ is an element of $GFL({\cal C})$.
\item[(GFL3)] If $\A\alw B \in GFL({\cal C})$ then $\A\nx\A\alw B \in GFL({\cal C}))$.
\item[(GFL4)] If $\E\ev B_{\sf ind} \in GFL({\cal C})$ then
$\E\nx\E\ev B_{\sf ind} \in GFL({\cal C})$.
\item[(GFL5)] If $\A\ev B \in GFL(G)$ then $\A\nx\A\ev B \in GFL({\cal C})$.
\item[(GFL6)] If $B \in GFL(G)$ and $B$ is not of the form $\lnot C$ then $\lnot B \in GFL({\cal C}))$.
\end{itemize}
\end{itemize}

Now given a set of ${\sf BNF}$ clauses represented by
$In \land \A\alw \phi$, we construct a labelled finite
graph ${\cal G}$ {\em incrementally} as follows

\begin{itemize}
\item[1.] The initial state is labelled by $In \land \A\alw \phi$.
\item[2.] Inductively assume that a graph has been constructed with
nodes labelled by the subsets of\\ $GFL(In \land \A\alw \phi)$,
where some formulae during such construction are marked.

{\em Note:} when providing the transitions from a state
$n$ if nodes $t_1$ and $t_2$ have the same label and the same marked
formulae, they are identified, and we delete one of them, say $t_2$, drawing an
edge to $t_1$. Also, if a transition leads to a node
which does not satisfy the propositional consistency criteria, we delete this node.

Given a node $n$ with the label $\Gamma$, choose an unmarked formula, say $B$, and apply an
appropriate expansion rule as follows:

\begin{itemize}
\item[2a.] If $B$ is an $\alpha$-formula, then create a successor of
$n$ labelled by\\ $\Gamma\union \{\alpha_1,\alpha_2\}$ and mark $B$ in the
label.

\item[2b.] If $B$ is a $\beta$-formula, then create two successors of
$n$ and label one of them by $\Gamma\union \{\beta_1\}$ while another one
by $\Gamma\union\{\beta_2\}$ and mark $B$ in the label.
\end{itemize}

\item[3.] If all non-elementary formulae within a node are marked,
such a node is called a {\em state}. Let $s$, be a state whose label
contains the following {\em next}-time formulae:
$$\A\nx B_1 \dots \A\nx B_k,\E\nx {C_1}_{\sf ind_1} \dots \E\nx {C_r}_{\sf ind_r}.$$
Merge all $C_i~(1 < i \leq q)$ which have identical indices,
(for example, given $\E\nx p_{\sf f}$ and  $\E\nx q_{\sf f}$,
producing $\E\nx (p\land q)_{\sf f}$) obtaining
in this way
$$\A\nx B_1 \dots \A\nx B_k, \E\nx{C_1}_{\sf ind_1}\dots \E\nx {C_m}_{\sf ind_m}.$$
Then create the successors $d_1\dots d_m$ of $s$ labelled respectively by
$$\{B_1 \dots  B_k, C_1\} \dots \{B_1 \dots  B_k, C_m\}.$$

\item[4.] Repeat steps 2 and 3 until no more new nodes are generated.
\end{itemize}

Now a tableau is a structure
${\cal G}_{{\cal C}_{Aug}} = (N, E, L)$, which satisfies the following conditions
\begin{itemize}
\item $N$ is a set of those nodes that are states in the construction above,
\item $E$ is a set of edges such that for $s,t\in N$, $E(s,t)$ if, and only if,
	$t$ is an immediate successor of $s$, and
\item $L$ is a set of labels, so for $s\in N$, the label of $s$ is $L(s)$.
\end{itemize}

In the following we will utilise the concept of pseudofulfillment of eventualities, adapting a
general definition to our case.

\begin{definition}
\label{ctl-pseudo-fulfilment-of-eventualities}
Given a tableau $_{{\cal C}_{Aug}}$, if a state $s$ contains
an eventuality $\eitherp\ev l$, then $\eitherp\ev l$ is
pseudo fulfilled, if ${\cal G}_R$ satisfies the following conditions.
	  \begin{itemize}
		\item If $\eitherp\ev l = \A\ev l$ then there
			  exists a finite subgraph $H$ of ${\cal G}_R$ with
			  $s$ as its root, such that for any terminal
			  state $t\in H$, $l\in t$.
		\item If $\eitherp\ev l = \E\ev l_{\sf f}$ then there
			  exists a finite subgraph $H$ of ${\cal G}_R$ with
			  $s$ as its root, such that $H$ has a finite
			  path $\pi_{s}$ which departs from $s$ and
		          satisfies the following condition: each state $t_{i+1}\in \pi_{s}$,
			  is the $f-th$ successor of $t_i$, and $l$ is
		          satisfied at the terminal state of $\pi_{s}$.
	  \end{itemize}
\end{definition}

Pseudo fulfilment informally means that for $\A\ev l$ constraints
we have a loop which contains a node satisfying  $l$ and for $\E\ev l_{\sf ind}$ constraints
we have a loop which contains a node satisfying $l$ where every successor node is an ${\sf ind}$-successor of the previous one, i.e. the one which is the successor node along the ${\sf ind}$ path.

Given a tableau ${\cal G}_{{\cal C}_{Aug}}$, apply the following deletion rules.
If a state has no successors, then delete this state and all
edges leading to it. If a state contains an eventuality which is not pseudo
fulfilled, delete this state.

Finally, if a state contains $\E\nx C_{\sf ind}$ and does not have
an ${\sf ind}$-successor which contains $C$, then delete this state. The resulting graph is called the reduced tableau.

\begin{theorem} (\cite{Bolotov-phd})
\label{emerson-tableau-decidability}
For any ${\sf BNF}$ set $\cal C$, its reduced tableau is empty, if and only if, $\cal C$ is unsatisfiable.
\end{theorem}

Now from a non-empty reduced graph ${\cal G}_{{\cal C}_{Aug}} = (N, E, L)$ for an augmented ${\sf BNF}$ we can construct
a B\"{u}chi Tree Automaton, ${\cal B} = \langle \Sigma,~D,~S,\delta,F_0,~F_B\rangle$ following the standard technique, for example,
\cite{Vardi06automata-theoretictechniques}.

\begin{itemize}
\item $\Sigma = 2^{Prop_{{\cal C}_{Aug}}}$, where $Prop_{{\cal C}_{Aug}}$ is a set of propositions of the clause set ${\cal C}_{Aug}$;
\item $D$ is the set of indices as defined for the construction of the Generalized Fischer-Ladner closure
in Def. \ref{ctl-fl-closure};
\item $S$ is a finite set, $N$, of states of ${\cal C}_{Aug}$;
\item $\delta$ is a transition function which corresponds to the edge relation $E$ of the reduced tableau;
\item $F_0$ is a set of those states that satisfy all the initial clauses
occurring within the clause set ${\cal C}_{Aug}$;
\item $F_B = S = N$ is a set of all states in the reduced tableau.
\end{itemize}

\begin{theorem}
\label{theorem-corectness-snf-to-buchi-automaton} Given a set,
${\cal C}$, of ${\sf BNF}$ clauses, we can construct a B\"{u}chi tree
automaton ${\cal B}$ such that ${\cal C}$ is satisfiable, if and
only if, ${\cal B}$ has an accepting run, $\tau_{\cal B}$.
\end{theorem}

{\bf Proof.} According to Theorem \ref{emerson-tableau-decidability}, if a set,
${\cal C}$, of ${\sf BNF}$ clauses is satisfiable then so is the reduced tableau for ${\cal C}_{Aug}$.
Due to the construction of the latter, following the transitions $E$, we can unwind it into an infinite tree $T$ such that
the root of the tree is labelled by the ${\cal C}_{Aug}$ and if a node $s_i\in N$ has ${\sf ind}$-successor nodes they become ${\sf ind}$-successors of the corresponding node $x_i$ of $T$.
The labelling of the states of $N$ gives the labelling of $\Sigma$ for the nodes of $T$. Now consider the $\Sigma$-labelled tree $T,\Sigma$ as a tree over which
the desired run $\tau$ of the automaton can be defined. The above construction of the automaton's accepting states guarantees that every path through the run would hit an accepting state infinitely often.

On the other hand, if we have an unsatisfiable set of ${\sf BNF}$ clauses
then its reduced tableau is empty, and therefore, the
automaton ${\cal B}$ whose construction is based upon the
properties of this reduced tableau, cannot have an accepting run. \cqd

\section{Example}
\label{example}

As an example of the syntactic representation of B\"{u}chi tree automaton in terms of ${\sf BNF}$ let us consider the following small automaton
${\cal B} = \langle \Sigma,~D,~S,\delta,F_0,~F_B\rangle$ where:
the alphabet $\Sigma = \{p,r\}$, the set of branching degrees is $D = \{d_1 = 2\}$; the set of states $S= \{s_0,s_1\}$ with the initial states $F_0 = s_0$ and
the accepting state $F_B = s_1$. Finally, let the transition function be defined as follows:

$\delta (s_0, p, 2) = <s_0,s_0>$, 
$\delta (s_1, p, 2) = <s_0,s_0>$, 
$\delta (s_0, r, 2) = <s_1,s_1>$, 
$\delta (s_1, r, 2) = <s_1,s_1>$.

This is a non-empty automaton, and its accepting run hits an accepting state  $s_1$ infinitely often on every branch. Now we will present
the components of the ${\sf BNF}$ for this automaton. With $q_1, q_2$, for the encoding of the states of the automaton and selecting $q_1$ to 
encode the initial state, in equations (\ref{eq:example-initial-states})-(\ref{eq:example-acceptance}) we give the components of the encoding of the given automaton, noting that $v$ in the representation of the labelling, and $y,l,u,w$ in the representations of the acceptance conditions are fresh variables.
First, the encoding of the initial states of the given automaton is represented by the set of clauses  (\ref{eq:example-initial-states}).

\begin{equation}
\label{eq:example-initial-states}
\begin{array}[t]{rl}
{\sf BNF}_{init_{\cal B}}&\\
\label{eq:example-initial-states}.1 &\start \imp  q_1\\
\end{array}
\end{equation}

Next, the transition function of the given automaton is represented by the set of clauses  (\ref{eq:example-transition-function}).

\begin{equation}
\label{eq:example-transition-function}
\begin{array}[t]{rcl}
{\sf BNF}_{tran_{\cal B}}&\\
\ref{eq:example-transition-function}.1 & q_1 \imp \E\nx {q_1}_{ind_ 1}\\
\ref{eq:example-transition-function}.2 & q_1 \imp \E\nx {q_2}_{ind_ 2}\\
\ref{eq:example-transition-function}.3 & q_2 \imp \E\nx {q_1}_{ind_ 1}\\
\ref{eq:example-transition-function}.4 & q_2 \imp \E\nx {q_2}_{ind_ 2}\\
\end{array}
\end{equation}

The set of clauses (\ref{eq:example-labelling}) represent the labelling of the given automaton.

\begin{equation}
\label{eq:example-labelling}
\begin{array}[t]{rcl}
{BNF}_{lab_{\cal B}}&\\
\ref{eq:example-labelling}.1 &\start &\imp \lnot q_1 \lor p \land r\\
\ref{eq:example-labelling}.2 &\start &\imp \lnot q_2 \lor p \land r\\
\ref{eq:example-labelling}.3 &\true  &\imp \A \nx (\lnot q_1 \lor v)\\
\ref{eq:example-labelling}.4 &     v &\imp p \land r\\
\ref{eq:example-labelling}.5 & \true  &\imp \A \nx (\lnot q_2 \lor v)\\
\end{array}
\end{equation}

Finally, for the automaton acceptance conditions we have the set of clauses  (\ref{eq:example-acceptance}), where $l, y, u, w$ are fresh variables.

\begin{equation}
\label{eq:example-acceptance}
\begin{array}[t]{rcl}
{BNF}_{acc_{\cal B}}&\\
\ref{eq:example-acceptance}.1 & \start &\imp y\\
\ref{eq:example-acceptance}.2 & \start &\imp \lnot l \lor q_1\\
\ref{eq:example-acceptance}.3 & \true &\imp \A\nx(\lnot l \lor q_1)\\
\ref{eq:example-acceptance}.4 & y   &\imp \A\alw u\\
\ref{eq:example-acceptance}.5 & u   &\imp \E\ev l_{ind_1}\\
\ref{eq:example-acceptance}.6 & l   &\imp \E\nx w_{ind_1}\\
\ref{eq:example-acceptance}.7 & w   &\imp \E\ev l_{ind_1}\\
\end{array}
\end{equation}


First, we show that the acceptance condition is properly simulated by the above ${\sf BNF}$. Indeed, every state along every path
of the underlying ${\sf BNF}$ computation tree, satisfies $u$, due to clauses \ref{eq:example-acceptance}.1 and \ref{eq:example-acceptance}.4. 
Now, pick an arbitrary state $s_i$ along some fullpath, say $\xi$. Following \ref{eq:example-acceptance}.5, there should be a state $s_j$, $i \leq j$, along the path
$Suf(\xi, s_j)$, which is labeled ${ind_1}$ such that it satisfies $l$. Following \ref{eq:example-acceptance}.6, the ${\sf ind}$-successor of the state of $s_j$, say $s_m$,
$j\leq m$ which satisfies $w$. Hence by \ref{eq:example-acceptance}.7, there should be a state, say, $s_k$, $m \leq k$ along the path $Suf(\xi, s_m)$ labeled by ${ind_1}$ such that $s_k$ satisfies $l$. Due to \ref{eq:example-acceptance}.3, states $s_j$ and $s_k$ satisfy $q_1$. Repeating this cain of reasoning steps regarding the satisfiability of $l$ we derive the recurrent satisfiability of $q_1$, which corresponds to the acceptance condition for this automaton, to hit the acceptance state $s_1$ infinitely often. 

\section{Discussion}
\label{discussion}

We have shown that normal form used for clausal resolution method for
a variety of CTL-type logics is expressive enough to give the succinct high-level syntactic
representation of B\"{u}chi tree automaton. This represents a significant
step in establishing the exact expressiveness of this formalism relating it to
this important class of tree automata. As a consequence, based on the known expressiveness results, that 
B\"{u}chi tree automata are as expressive as CTL extended by the propositional quantification, we can also 
treat {\sf BNF} as a kind of normal form for the latter: we can directly translate given problem specifications into ${\sf BNF}$
and apply as a verification method a {\em deductive reasoning technique} -- the temporal resolution technique.
This paper justifies that ${\sf BNF}$ is suitable to reason about a wider range of branching-time specifications and is able to capture exactly those that are captured by tree automata.

Moreover, another very promising route is emerging here. In ${\sf BNF}$ we are enabled not only to represent 
some given, or explicit invariants, but also to discover implicit, hidden invariants.
For example, invariants are extremely important in the emerging trend of developing complex but re-configurable software systems. Here one of the most challenging problems is assuring that invariants are maintained during system reconfiguration. In \cite{pltl-invariants-fisher}, it has been shown how the clausal temporal resolution technique developed for temporal logic provides an effective method for searching for invariants in the linear time setting. 
The results of this paper enable the extension of this method to branching-time framework, 
and the detailed analysis of this route forms one direction of future research. 

Also, in the future we will investigate the representation of alternating tree automata in ${\sf BNF}$, where we expect results similar to the linear-time case \cite{BFD02}.

\nocite{*}
\bibliographystyle{eptcs}
\bibliography{biblio}
\end{document}